\documentclass[twocolumn, nofootinbib, superscriptaddress, showkeys]{revtex4-2}
\usepackage{header}

\begin{document}
\title{Extended spatial coherence of interlayer excitons in MoSe$_2$/WSe$_2$ heterobilayers}
\date{\today}
\def\wsi{Walter Schottky Institute and Physics Department, TU Munich, Am Coulombwall 4a, 85748 Garching, Germany}
\def\mcqst{Munich Center for Quantum Science and Technology (MCQST), Schellingstr. 4, 80799 Munich, Germany}
\def\berlin{Institute for Theoretical Physics, Nonlinear Optics and Quantum Electronics, Technical University of Berlin, 10623 Berlin, Germany}
\def\muenster{Institute of Physics, Münster University, Wilhelm-Klemm-Str. 10, 48149 Münster, Germany}

\author{Mirco Troue}\thanks{Both authors contributed equally.}
\author{Johannes Figueiredo}\thanks{Both authors contributed equally.}

\author{Lukas Sigl}
\author{Christos Paspalides}\affiliation{\wsi}\affiliation{\mcqst}

\author{Manuel Katzer}\affiliation{\berlin}

\author{Takashi Taniguchi}
\affiliation{International Center for Materials Nanoarchitectonics, National Institute for Materials Science, Tsukuba 305-0044, Japan}
\author{Kenji Watanabe}
\affiliation{Research Center for Functional Materials, National Institute for Materials Science, Tsukuba 305-0044, Japan}

\author{Malte Selig}
\author{Andreas Knorr}\affiliation{\berlin}

\author{Ursula Wurstbauer}\affiliation{\muenster}

\author{Alexander W. Holleitner}\email[Electronic address: ]{holleitner@wsi.tum.de}\affiliation{\wsi}\affiliation{\mcqst}
\begin{abstract}
    We report on the spatial coherence of interlayer exciton ensembles as formed in \ce{MoSe2}/\ce{WSe2} heterostructures and characterized by point-inversion Michelson-Morley interferometry. Below \SI{10}{\kelvin}, the measured spatial coherence length of the interlayer excitons reaches values equivalent to the lateral expansion of the exciton ensembles. In this regime, the light emission of the excitons turns out to be homogeneously broadened in energy with a high temporal coherence. At higher temperatures, both the spatial coherence length and the temporal coherence time decrease, most likely because of thermal processes. The presented findings point towards a spatially extended, coherent many-body state of interlayer excitons at low temperature.
\end{abstract}

\keywords{Many-body state, quantum coherence, interlayer excitons, two-dimensional materials }

\maketitle
In the rapidly growing family of two-dimensional materials, many-body excitations of charge carriers play an essential role to describe the emergent quantum phenomena in the atomistic materials with ultimately confined dimensions in one direction and a low dielectric screening \cite{Wilson2021}. Particularly, van der Waals heterostructures of semiconducting two-dimensional materials, such as \ce{MoSe2} and \ce{WSe2}, are ideal systems to study the many-body phase diagram of interlayer excitons (IXs) \cite{Blatt1962a, Fogler2014a}. The latter are Coulomb-bound electron-hole pairs, where the photo-generated electron is localized in one layer and the hole in the other layer of a heterostructure. Such excitons are composite bosons and turn out to exhibit a large exciton binding energy of several hundreds of \si{\milli\electronvolt}, long photoluminescence lifetimes exceeding the thermalization timescales, and a possible gate-tunable interlayer hybridization \cite{Rivera2015a, Miller2017a, Hanbicki2018a, Merkl2019, Kiemle2020a}. In order to experimentally verify the emergence of a spatially extended quantum many-body state, the exciton ensembles have to comprise mobile particles such that each exciton can interact, e.g. via a dipole-dipole repulsion with other excitons in the ensemble \cite{Blatt1962a, Fogler2014a, Laikhtman2009a, Griffin1995}. The corresponding free expansion of excitons has been proven for interlayer excitons in several geometries and heterostructures \cite{Gartner2006, Vogele2009a, Wagner2021, Sun2022, Peng2022, Datta2022}. Recently, it was demonstrated that interlayer excitons can show an enhanced temporal coherence time, i.e. a reduced spectral linewidth in the predicted degeneracy limit where the excitons are supposed to coherently interact with each other \cite{Sigl2020a}. However, it was pointed out in several reports on many-body exciton ensembles \cite{Sigl2020a, Eisenstein2004a, High2012, Alloing2014, Stern2014, Liu2017, Kogar2017a, Wang2019b}, that it is required to demonstrate the extended spatial coherence of the excitons as an evidence of a coherent many-body state. In this letter, we demonstrate that the spatial coherence length of interlayer excitons in van der Waals \ce{MoSe2}/\ce{WSe2}-heterostructures can reach values equal to the spatial expansion length of the exciton ensembles, when they are measured in the few Kelvin regime. Our results open the pathway for future quantum technological devices based on coherent exciton ensembles with the possibility to integrate them in laterally patterned, logical circuits \cite{Schinner2013a, Shanks2021a}.

\begin{figure}
    \centering
    \includegraphics{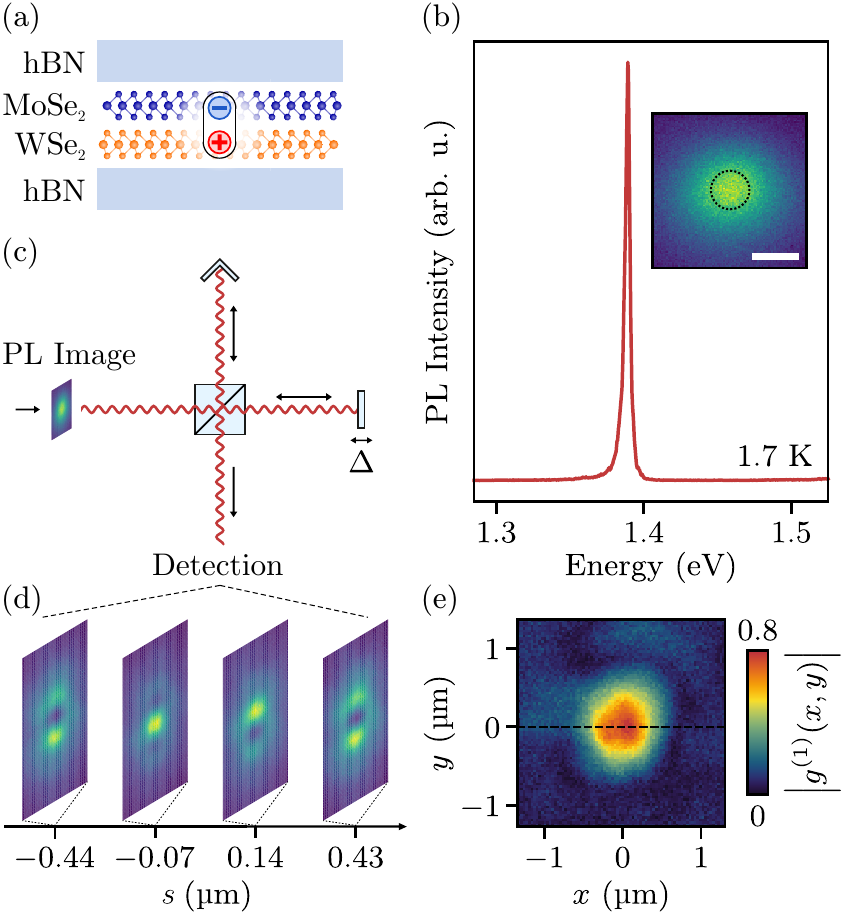}
    \caption{Spatial coherence of interlayer excitons in a \ce{MoSe2}/\ce{WSe2} heterobilayer. (a) Sketch of a \ce{MoSe2}/\ce{WSe2} heterobilayer encapsulated in \ce{hBN}. After optical excitation, interlayer excitons (IXs) form with the electron (hole) in the \ce{MoSe2} (\ce{WSe2}). (b) Typical photoluminescence spectrum of such IXs with an emission energy of $E_{\text{IX}} = \SI{1.389}{\electronvolt}$ ($\lambda_{\text{IX}} = \SI{892.6}{\nano\meter}$) at a bath temperature of $T_{\text{bath}} = \SI{1.7}{\kelvin}$ ($E_{\text{laser}} = \SI{1.94}{\electronvolt}$ and $P_{\text{laser}} = \SI{2}{\micro\watt}$). The inset shows a corresponding spatially resolved photoluminescence image of the sample at $P_{\text{laser}} = \SI{3}{\micro\watt}$. Scale bar, \SI{1}{\micro\meter}. Dotted circle denotes the extent of the measured point spread function (PSF) of the optical circuitry at $\lambda_{\text{IX}}$; measured as a convolution of the excitation- and detection-path at this wavelength. (c) Concept of the point-inversion Michelson-Morley interferometry. A central beam splitter distributes the spatially resolved photoluminescence image of the sample into the two arms of the interferometer. In one of the arms, the image is point-inverted by the help of a retro-reflector, while in the other, the image is reflected by a mirror, before the interference image is detected. A piezo-stepper attached to the mirror sets the optical path difference $s$ between both arms. (d) Four exemplary interference images of the photoluminescence image as in the inset of (a) vs. the path difference $s$. (e) Corresponding spatially resolved coherence image $|g^{(1)}(x, y)|$ at $T_{\text{bath}} = \SI{1.7}{\kelvin}$.}
    \label{fig:fig1}
\end{figure}

The investigated samples consist of stacked heterobilayers of \ce{MoSe2} and \ce{WSe2} monolayers in an H-type configuration, that are encapsulated in hexagonal boron nitride (\ce{hBN}) [\cref{fig:fig1}(a)]. The corresponding photoluminescence spectra show a single Lorentzian emission peak at $E_{\text{IX}} = \SI{1.389}{\electronvolt}$ ($\lambda_{\text{IX}} = \SI{892.6}{\nano\meter}$) at low excitation powers and low temperatures [\cref{fig:fig1}(b)], as consistent with earlier reports \cite{Sigl2020a, Sigl2022}. The luminescent peak can be interpreted to stem from interlayer excitons where the electron (hole) is localized in the \ce{MoSe2} (\ce{WSe2}). The narrow Lorentzian emission line persists down to the lowest investigated excitation powers without separating into distinct lines, and it features a lifetime of several tens of \si{\nano\second}. In turn, we interpret the interlayer excitons to occupy the lowest excitonic energy state of the corresponding composite-boson ensemble. The inset in \cref{fig:fig1}(b) shows a spatial image of the excitonic photoluminescence $I(x,y)$ as a function of the coordinates $x$ and $y$. Clear photoluminescence signal is detected several hundreds of nanometers beyond the measured point spread function (PSF) at $\lambda_{\text{IX}}$, which amounts to \SI{899(18)}{\nano\meter} in the utilized optical circuitry (cf. dotted circle). As a result, we infer that the imaged long-lived interlayer excitons propagate hundreds of nanometers within the plane of the studied heterobilayer \cite{Gartner2006, Vogele2009a, Wagner2021, Sun2022, Peng2022, Datta2022}.

\begin{figure}
    \centering
    \includegraphics{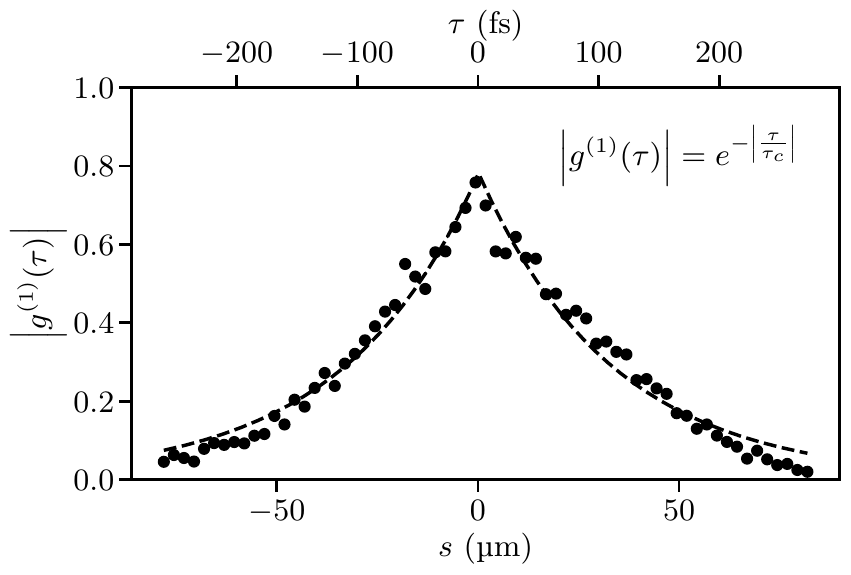}
    \caption{Temporal coherence of the excitons. Normalized value of the first order correlation function $|g^{(1)}(\tau)| \equiv |g^{(1)}(\tau, x = 0, y = 0)|$ as in \cref{fig:fig1}(e) for $x, y = 0$ but as a function of $s$ and therefore, $\tau = s / c$, with $c$ the speed of light. All experimental parameters are as in \cref{fig:fig1}(e) with $T_{\text{bath}} = \SI{1.7}{\kelvin}$. The full width at half maximum (FWHM) of $g^{(1)}(\tau)$ gives the temporal coherence time $\tau_\text{c} = \SI{223(6)}{\femto\second}$, consistent with results on similar samples as in ref. \cite{Sigl2020a}. The two exponential curves fit the data as expected for a purely homogeneously broadened light emission.}
    \label{fig:fig2}
\end{figure}

\begin{figure*}
    \centering
    \includegraphics{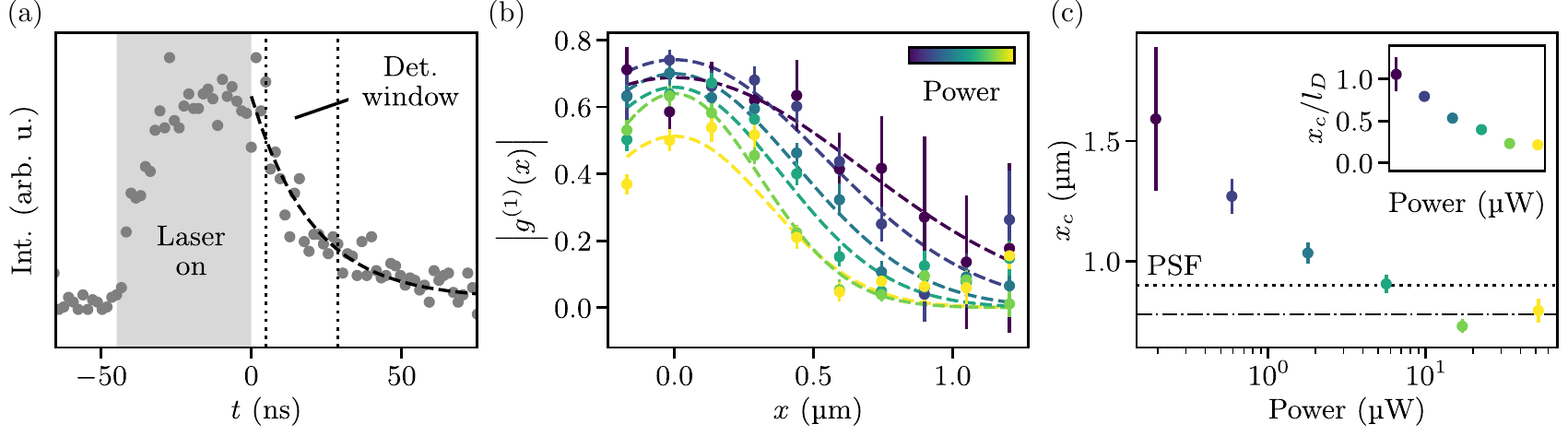}
    \caption{Thermalized excitons and their spatial coherence. (a) Time-resolved photoluminescence of interlayer excitons at $E_\text{IX} = \SI{1.389}{\electronvolt}$. After the laser is turned off, the photoluminescence is fitted by an exponential decay with a lifetime of \SI{20.3(13)}{\nano\second} (dashed line). For the interference experiments, the signal is detected during a time window of \SI{22.4}{\nano\second} at a time delay of \SI{6.4}{\nano\second} after the laser is turned off. The chosen time delay ensures that the IXs are thermalized and that laser-induced coherence effects are excluded during the light-detection. (b) $|g^{(1)}(x)|$ of the signal within the detection window of (a) for several laser powers as a function of the spatial coordinate $x$, i.e. $|g^{(1)}(x)| \equiv |g^{(1)}(\tau = 0, x, y = 0)|$ as highlighted by a dashed line in \cref{fig:fig1}(e). The maximum spatial coherence occurs at the lowest power ($P_\text{laser} = \SI{200}{\nano\watt}$), while $|g^{(1)}(x)|$ is reduced for higher laser intensities. Dashed lines are Gaussian fits to the data to extract the full width at half maximum (FWHM) of the spatial distributions, which is the spatial coherence length $x_\text{c}$. (c) Dependence of $x_\text{c}$ vs. $P_\text{laser}$. At the lowest presented power, $x_\text{c}$ exceeds the PSF of the optical circuitry, while for high powers, $x_\text{c}$ is consistent with the PSF as expected for a transition from a spatially coherent ensemble to an incoherent ensemble of IXs. Dotted line: PSF at $\lambda_\text{IX}$. Dotted-dashed line: PSF measured at the excitation wavelength. Inset: ratio of $x_\text{c}$ to the full width at half maximum $l_\text{D}$ of the overall spatial photoluminescence images [cf. inset of \cref{fig:fig1}(b)].}
    \label{fig:fig3}
\end{figure*}

We perform a point-inversion Michelson-Morley interferometry to characterize the spatial and temporal coherence of the excitonic photoluminescence. Starting point are spatially homogeneous IX photoluminescence images as in the inset of \cref{fig:fig1}(b). In the interferometer, a centered 50/50 beam splitter distributes the incoming photoluminescence image equally into the two arms of the interferometer [\cref{fig:fig1}(c)]. In one of the arms, a three-sided retroreflector point-inverts the image, i.e. $I_1(x, y) \rightarrow I_1(-x,-y)$, while in the second arm, a plane mirror reflects the image, $I_2(x,y) \rightarrow I_2(x,y)$. Both the point-inverted and the unchanged image pass the beam splitter again and interfere with each other in the detection path. The superposition is finally detected and reveals the spatial interference relative to the inversion center \cite{Paik2019a}. We utilize a piezo stepper to move the mirror in the second arm by a distance $\Delta$, which introduces an optical path difference of $s = 2\Delta$ and a relative time delay $\tau = s / c$ between the two paths of the interferometer arms (with $c$ the speed of light). For each image position, we observe interference fringes with a period consistent with the exciton emission wavelength $\lambda_{\text{IX}}$. \cref{fig:fig1}(d) shows such interference images for $s = \SI{-0.44}{\micro\meter}$, \SI{-0.07}{\micro\meter}, \SI{0.14}{\micro\meter}, and \SI{0.43}{\micro\meter}. One can clearly see how the spatial interference patterns vary with $s$ and therefore, with the time-delay $\tau$. Per $x$- and $y$-coordinate of the original image, the detected intensity can be described as \cite{Loudon2000}:
\begin{eqnarray}
    I\left(\tau, x, y\right) = I_1 + I_2 + & 2\sqrt{I_1 \left(0,-x,-y \right) I_2 \left(\tau,x,y \right)} \nonumber \\ & g^{(1)}(\tau,x,y) \cos{\left(\dfrac{2\pi}{\lambda_{\text{IX}}} c\tau \right)}\,,
    \label{eq:eq1}
\end{eqnarray}
with $g^{(1)}(\tau, x, y)$ the normalized first-order correlation function given by
\begin{equation}
    g^{(1)}\left( \tau, x, y \right) = \dfrac{\left\langle E(0,-x,-y) E^*(\tau,x,y) \right\rangle}{\sqrt{I_1(0,-x,-y) I_2(\tau,x,y)}}
    \label{eq:eq2}
\end{equation}
and $E = E(\tau, x, y)$ the electric field and $E^* = E^*(\tau, x, y)$ its complex conjugate of the interfering photons. In other words, images such as in \cref{fig:fig1}(d) allow to determine $|g^{(1)}(\tau, x, y)|$, which is proportional to the visibility of the interference. The later is limited by the ratio of the light intensities in the two arms of the interferometer to $2 \tfrac{\sqrt{I_1 I_2}}{I_1+I_2}$, which we calculate to be \num{0.98} in our interferometer. The limit stems from the fact that the retroreflector in the first arm of the interferometer has a slightly reduced total reflectance in comparison to the mirror in the second arm. All presented values of $|g^{(1)}(\tau, x, y)|$ are normalized to this technical limit.
In a first step, we characterize the temporal coherence of the excitonic photoluminescence in the center of the excitation spot. The first-order correlation function implies the temporal coherence time of the excitonic photoluminescence, when the time-delay $\tau$  is varied for $x = y = 0$. Namely, a variation of the optical path difference in the center of the PSF, i.e. the central pixel of the detection spot, introduces the necessary temporal offset between the interfering photoluminescence images. We write $|g^{(1)}(\tau)| \equiv |g^{(1)}(\tau, x = 0, y = 0)|$ and plot $|g^{(1)}(\tau)|$ as a function of $s$ and therefore $\tau$ (\cref{fig:fig2}). As expected, for $\tau = 0$, $|g^{(1)}(\tau)|$ reaches the maximum value as in \cref{fig:fig1}(e). Fitting the data with an exponential fit (dashed lines), we determine the maximum to be $|g^{(1)}(\tau = 0)| = \num{0.76(2)}$ ($= \SI{76}{\percent}$). We note that for an excitation at lower power, a value of \SI{88}{\percent} is observed, as discussed below. The full width at half maximum (FWHM) of the data in \cref{fig:fig2} is $\tau_\text{c} = \SI{233(6)}{\femto\second}$, which is the so-called temporal coherence time of the excitonic photoluminescence. We point out that the Lorentzian shape of the investigated photoluminescence spectra, as in \cref{fig:fig1}(b), warrants the use of the exponential fit in \cref{fig:fig2} \cite{Loudon2000}. For the particular spectrum underlying the measurement of \cref{fig:fig2}, the energetic photoluminescence linewidth is \SI{5.60(4)}{\milli\electronvolt}. In turn, a temporal coherence time of \SI{235(17)}{\femto\second} is expected, which agrees with the experimentally determined value of $\tau_\text{c}$ at the same power and within the given error. In turn, the excitonic photoluminescence such as in \cref{fig:fig1}(b) fulfills the Wiener-Khinchin theorem for a homogeneously broadened emitter \cite{Loudon2000}. We note that our finding on the temporal coherence time of interlayer excitons is consistent with an earlier report on similar \ce{MoSe2}/\ce{WSe2}-heterobilayers \cite{Sigl2020a}.
In general, the coherence of exciton ensembles is affected by local heating effects due to photoexcitation and the impact of both hot charge carriers and extra excitons. Moreover, a continuous laser excitation permanently induces coherence and enhances the interferometric visibility of the exciton photoluminescence. In order to minimize such extrinsic effects, we use a pulsed laser excitation and detect the excitonic photoluminescence images of the interlayer excitons at a delayed time, where the laser is sufficiently long turned off. \Cref{fig:fig3}(a) depicts the total emission of a photoluminescence image, such as the one in the inset of \cref{fig:fig1}(b), as a function of the time delay after the laser excitation. The image is spectrally integrated in the range $\SI{1.378}{\electronvolt} \leq E_{\text{photon}} \leq \SI{1.442}{\electronvolt}$, with $E_{\text{photon}}$ the photoluminescence energy, such that only the exciton emission at $E_{\text{IX}} = \SI{1.389}{\electronvolt}$ is detected but no other optical transition for all laser powers shown. The dashed line in \cref{fig:fig3}(a) is an exponential fit to the data giving the photoluminescence lifetime to be \SI{20.3(13)}{\nano\second} for the specific experimental parameters. For the spatial coherence measurements, \cref{fig:fig3}(b), we choose the detection window with a width of \SI{22.4}{\nano\second} to start \SI{6.4}{\nano\second} after the laser pulse; an offset time which exceeds the thermalization and decay time of hot charge carriers and intralayer excitons in the femto- to picosecond regime \cite{Fang2019a}. The chosen offset time also exceeds the decay time of interlayer excitons at higher emission energies, which are in the single nanosecond regime \cite{Rivera2015a, Sigl2020a}. The laser-induced coherence can be excluded as well. Within the detection window [dotted lines in \cref{fig:fig3}(a)], we determine $|g^{(1)}(\tau = 0, x, y = 0)| \equiv |g^{(1)}(x)|$, i.e. the modulus of the first order auto correlation function along the $x$-coordinate of the excitonic photoluminescence images [e.g. dashed line in \cref{fig:fig1}(e)]. To do so, the optical path difference $\tau$ is symmetrically varied around zero: $\SI{-1.67}{\femto\second} \leq \tau \leq \SI{+1.67}{\femto\second}$ ($\SI{-0.5}{\micro\meter} \leq s \leq \SI{0.5}{\micro\meter}$) and corresponding interference images are analyzed. We note that the already introduced width of the temporal detection window of \SI{22.4}{\nano\second} is chosen in order to achieve a high enough signal-to-noise ratio for this analysis. \cref{fig:fig3}(b) depicts the corresponding $|g^{(1)}(x)|$ for several laser powers. For each laser power, we fit $|g^{(1)}(x)|$ with a Gaussian function [dashed lines in \cref{fig:fig3}(b)] to reveal the full width at half maximum, which is the spatial coherence length $x_\text{c}$ of the excitonic emission. \Cref{fig:fig3}(c) compares $x_\text{c}$ to the PSF for several laser powers. For $P_{\text{laser}} = \SI{200}{\nano\watt}$, we measure a spatial coherence length of $x_\text{c} = \SI{1.6(3)}{\micro\meter}$, which significantly exceeds the measured PSF (dotted line), while for high excitation powers, the data are well described by the measured PSF of the optical circuitry. We tentatively explain the decrease of the spatial coherence for high laser powers by the increasing influence of exciton-exciton interactions, e.g. with excitons at higher kinetic energies in the exciton ensembles \cite{Katsch2020}. For even higher powers, further exciton transitions at higher emission energies eventually start to dominate the photoluminescence spectra (data not shown, since already discussed in \cite{Sigl2020a} on other samples). In the presented figures, however, all powers are below this transition, and all corresponding photoluminescence spectra exhibit only one emission line as in \cref{fig:fig1}(b).
In order to compare the spatial coherence length $x_\text{c}$ to the lateral expansion of the interlayer excitons, we first determine the spatial expansion of the excitons by fitting the spatial photoluminescence images, such as in the inset of \cref{fig:fig1}(b), with two-dimensional Gaussian distributions and define their isotropic FWHM as the excitonic expansion length $l_\text{D}$. Plotting the ratio of $x_\text{c}$ vs. $l_\text{D}$ [inset of \cref{fig:fig3}(c)], we observe that for the lowest investigated power of \SI{200}{\nano\watt}, the spatial coherence length reaches the value of the spatial expansion of the exciton ensemble within the plane of the \ce{MoSe2}/\ce{WSe2}-heterobilayer. For high laser powers, the ratio decreases to a value of $\approx \num{0.2}$. The decrease can be explained by a transition from a spatially coherent exciton ensemble gradually transitioning into an incoherent ensemble; namely the observed value is explained by the ratio of the PSF to the corresponding expansion of the excitonic ensemble.

\begin{figure}
    \centering
    \includegraphics{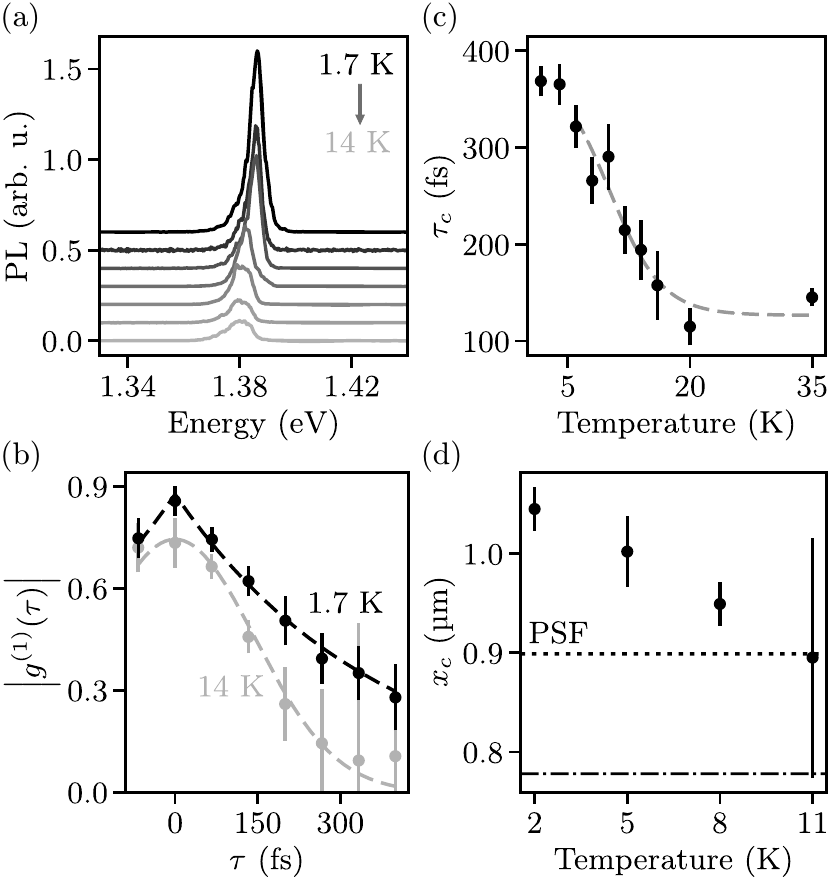}
    \caption{Temperature dependence of the temporal and spatial coherence lengths. (a) Photoluminescence spectra of IXs for $T_\text{bath} = 1.7, 4, 6, \dots, 14 \si{\kelvin}$. The emission amplitudes decrease with increasing $T_\text{bath}$. (b) $|g(1)(\tau)|$ for $T_{\text{bath}} = \SI{14}{\kelvin}$ (gray) and \SI{1.7}{\kelvin} (black). (c) Temporal coherence time $\tau_\text{c}$ vs. $T_\text{bath}$. Gray dashed line is a guide to the eye. (d) Spatial coherence length $x_\text{c}$ vs. $T_\text{bath}$. Time delay: \SI{27.2}{\nano\second} and time window of \SI{16}{\nano\second}. Dotted line (dotted-dashed line) is the PSF at $\lambda_\text{IX}$ (at the excitation wavelength). Experimental parameters for all subfigures are $E_{\text{laser}} = \SI{1.94}{\electronvolt}$ and $P_{\text{laser}} = \SI{400}{\nano\watt}$.}
    \label{fig:fig4}
\end{figure}

In a next step, we discuss the temperature dependence of the exciton dynamics. \cref{fig:fig4}(a) depicts the photoluminescence spectra for an increasing bath temperature. Above \SI{10}{\kelvin}, the intensity decreases significantly, while $|g^{(1)}(\tau)|$ deviates from the Lorentzian profile approaching a Gaussian one, and it is not described by double-exponential curves as for below \SI{10}{\kelvin} [\cref{fig:fig4}(b)]. We note that for the data of \cref{fig:fig4}(b), $|g^{(1)}(\tau)|$ reaches \num{0.88(3)} (\SI{88}{\percent}). The apparent difference to \cref{fig:fig2} can be explained by a lower excitation power and in turn, by a reduced impact of excess energy at the excitation spot (\SI{3}{\micro\watt} in \cref{fig:fig2} vs. \SI{400}{\nano\watt} in \cref{fig:fig4}(b)). \cref{fig:fig4}(c) shows the extracted temporal coherence time $\tau_\text{c}$ vs. bath temperature at this lower power. For the lowest temperatures, $\tau_\text{c}$ reaches a value of up to \SI{370}{\femto\second}. We detect this maximum when the laser is ``on'' and again, for a low laser power. The maximum value monotonously decreases for the detection being delayed with respect to the laser irradiation (data not shown). The gray dashed line in \cref{fig:fig4}(c) is a guide to the eye. Last but not least, \cref{fig:fig4}(d) depicts the spatial coherence length $x_\text{c}$ vs. the bath temperature, again analyzed for a detection window when the laser is ``off''. Within the given uncertainty of the interferometric method, $x_\text{c}$ decreases and approximates the PSF for high temperatures, as will be discussed in the following.

Generally, the investigated ensembles of interlayer excitons exhibit spatially homogeneous photoluminescence images [inset of \cref{fig:fig1}(b)], where the photon emission exceeds the spatial extent of the measured PSF by several hundreds of nanometers. As a consequence, we infer that at least, a sub-ensemble of the interlayer excitons is mobile \cite{Gartner2006, Vogele2009a, Wagner2021, Sun2022, Peng2022, Datta2022}. Significantly, the spatial coherence length matches the expansion length at a low excitation power [inset of \cref{fig:fig3}(c)]. In turn, localization phenomena do not seem to hinder the presented spatial coherence phenomena in the investigated regime \cite{Lagoin2021a, Bieniek2022}. We estimate the exciton density to be on the order of $\approx \SI{1e11}{\per\centi\meter\squared}$ suggesting that the exciton ensembles are degenerate below \SI{10}{\kelvin} \cite{Sigl2020a}; namely that the excitonic thermal de Broglie wavelength of several tens of nanometers exceeds the inter-excitonic distance in this temperature range. As a result, we assume the underlying exciton expansion to be a quantum mechanical percolation process within the two-dimensional potential landscape of the heterobilayer \cite{Glazov2022}. As far as the temporal coherence time is concerned, its maximum reaches a value of $\tau_\text{c} \approx \SI{370}{\femto\second}$ at low temperature [\cref{fig:fig4}(c)], as it is consistent with \cite{Sigl2020a}. Tentatively, we interpret this fast process by a possible momentum and energy transfer during the photon-emission process \cite{Lagoin2021a, Remez2022}. For low temperatures, a coupling to an underlying low-lying collective excitation of the exciton ensembles is reasonable \cite{Remez2022, Dietl2017}, while for higher temperatures, a thermally activated momentum transfer is very likely. Consistently, above \SI{10}{\kelvin}, we observe that $|g^{(1)}(\tau)|$ can be fitted by Gaussian curves expressing an inhomogeneous broadening of the photoluminescence [\cref{fig:fig4}(b)] \cite{Loudon2000}. Moreover, we observe that the temporal coherence time decreases [\cref{fig:fig4}(c)]. Both findings suggest that the excitons are in a thermal regime with significantly reduced coherence at elevated temperatures. \cref{fig:fig4}(d) shows the spatial coherence length $x_\text{c}$ vs. bath temperature. We restrict ourselves to the temperature range of $\SI{1.7}{\kelvin} \leq T_\text{bath} \leq \SI{11}{\kelvin}$, because above \SI{10}{\kelvin}, the photoluminescence amplitude significantly drops [\cref{fig:fig4}(a)] and as already discussed, the light emission is more and more inhomogeneously broadened. For $\SI{1.7}{\kelvin} \leq T_\text{bath} \leq \SI{11}{\kelvin}$, the spatial coherence length decreases [\cref{fig:fig4}(d)] most likely because of the increased influence of thermally activated processes. At higher temperatures, the noise appearing for $x_\text{c}$ increases because of the reduced photoluminescence amplitude [cf. \cref{fig:fig4}(a)], but the PSF is the natural limit of the spatial coherence for higher temperatures [\cref{fig:fig4}(d)]. So far, we compared the data to the PSF at $\lambda_\text{IX}$ which amounts to \SI{899(18)}{\nano\meter} [dotted lines in \cref{fig:fig1}(b), \cref{fig:fig3}(c), and \cref{fig:fig4}(d)], measured as a convolution of the excitation- and detection-path at this wavelength. The similarly measured PSF at the utilized excitation wavelength of \SI{639}{\nano\meter} is \SI{779(18)}{\nano\meter} [dashed-dotted line in \cref{fig:fig3}(c) and \cref{fig:fig4}(d)]. In turn, we can expect that the excitation spot is on the order of \SI{550}{\nano\meter} ($\approx \SI{779}{\nano\meter} / \sqrt{2}$). Consistently, the data in the presumably incoherent regime at high laser power and/or high bath temperatures lie in-between the two mentioned PSFs [dotted and dashed-dotted lines in \cref{fig:fig3}(c) and \cref{fig:fig4}(d)].

Future studies might resolve the transition to single excitons at even lower excitation powers, as was done for interlayer excitons in III-V semiconductor double quantum wells \cite{Schinner2013a}, as well as the impact of exciton-exciton and exciton-phonon interactions on the spatial coherence at higher exciton densities and/or lower temperatures in combination with the bosonic and fermionic aspects of interactions within the exciton ensembles \cite{Bieniek2022}. We point out that the high interference visibility of the interlayer exciton ensembles is comparable to the one reported for exciton-polaritons in nanophotonic devices based on similar heterobilayers ($\approx \SI{80}{\percent}$ in \cite{Paik2019a}). In turn, our results suggest the general feasibility of future quantum devices based on spatially confined coherent exciton ensembles interacting coherently with photons \cite{Vogele2009a, Peng2022, Shanks2021a}. Last but not least, we note that for an experimental evidence of an exciton condensation in the momentum space, a back-focal plane imaging experiment seems to be suitable, however, at bath temperatures significantly below \SI{1}{\kelvin} \cite{Sigl2022}, which is beyond the scope of the current study.
In summary, we show that the photoluminescence of interlayer excitons, as formed in hetero­bilayers of \ce{MoSe2}- and \ce{WSe2}-monolayers can exhibit an extended spatial coherence length exceeding the PSF of the utilized optical circuitry at bath temperatures below \SI{10}{\kelvin}. The spatial coherence length reaches the spatial extent of the exciton ensemble in the investigated regime, and it decreases to the PSF for increasing laser intensities and/or bath temperatures. The photoluminescence of the coherent exciton ensemble turns out to exhibit a temporal coherence time of several 100s of femto­seconds and a high interference visibility. We determine the temperature dependence of both the temporal coherence time and the spatial coherence length, suggesting that above \SI{10}{\kelvin} thermal processes start to dominate the exciton interactions fostering the transition to a non-degenerate exciton gas.

\section*{Competing Interests}
The authors declare no competing interest nor conflict.

\begin{acknowledgments}
    The authors gratefully acknowledge the German Science Foundation (DFG) for financial support via grants HO 3324/9-2 and WU 637/4-2 and 7-1, as well as the clusters of excellence MCQST (EXS-2111) and e-conversion (EXS-2089) as well as the priority program 2244 (2DMP). K.W. and T.T. acknowledge support from JSPS KAKENHI (Grant Numbers 19H05790, 20H00354 and 21H05233).\\\\\\
\end{acknowledgments}
\bibliography{bib}
\end{document}